\documentclass[12pt]{article}

\usepackage{latexsym}

\usepackage{epsfig}
\usepackage{psfrag}

\def\void{}
\def\labelmark{}

\newenvironment{formula}[1]{\def\labelname{#1}
\ifx\void\labelname\def\junk{\begin{displaymath}}
\else\def\junk{\begin{equation}\label{\labelname}}\fi\junk}%
{\ifx\void\labelname\def\junk{\end{displaymath}}
\else\def\junk{\end{equation}}\fi\junk\labelmark\def\labelname{}}

{\ifx\void\labelname\def\junk{\end{array}\end{displaymath}}
\else\def\junk{\end{array}\right.\end{equation}}
\fi\junk\labelmark\def\labelname{}\def\junk{}
}

\newcommand{\beq}{\begin{formula}}
\newcommand{\eeq}{\end{formula}}
\newcommand{\beqv}{\begin{formula}{}}

\newcommand{\rf}[1]{(\ref{#1})}
\newcommand{\oh}{\frac{1}{2}}

\newcommand{\bea}{\begin{eqnarray}}
\newcommand{\eea}{\end{eqnarray}}
\newcommand{\beas}{\begin{eqnarray*}}
\newcommand{\eeas}{\end{eqnarray*}}
\newcommand{\beqs}{\begin{displaymath}}
\newcommand{\eeqs}{\end{displaymath}}

\newcommand{\br}{\langle}
\newcommand{\kt}{\rangle}
\newcommand{\ep}{\epsilon}

\newcommand{\vp}{\varphi}
\newcommand{\Tr}{{\rm Tr}\;}
\newcommand{\ben}{\begin{equation}}
\newcommand{\een}{\end{equation}}

\newcommand{\bdm}{\begin{displaymath}}
\newcommand{\edm}{\end{displaymath}}
\newcommand{\bR}{{\bf R}}

\begin{document}
 \topmargin 0pt
 \oddsidemargin 5mm
 \headheight 0pt
 \topskip 0mm

 \addtolength{\baselineskip}{0.4\baselineskip}

 \pagestyle{empty}

 \vspace{0.5cm}

\hfill RH-02-2001

\vspace{2cm}

\begin{center}

{\Large \bf The Existence and Stability of \\ Noncommutative Scalar Solitons}

\medskip

\vspace{.5 truecm}

%\centerline{}

%\vspace{.2 truecm}

%\vspace{1.5cm}

 \vspace{0.7 truecm}
Bergfinnur Durhuus$^a$\footnote{email: durhuus@math.ku.dk}, Thordur 
Jonsson$^b$\footnote{e-mail: thjons@raunvis.hi.is}
and Ryszard Nest$^a$\footnote{email: rnest@math.ku.dk}

\vspace{.5 truecm}

$^a$Matematisk Institut, Universitetsparken 5

2100 Copenhagen \O, Denmark

 \vspace{.5 truecm}

$^b$University of Iceland, Dunhaga 3,

107 Reykjavik, Iceland

% \vspace{1.5 truecm}

 \vspace{.5 truecm}

 \end{center}

 \noindent
 {\bf Abstract.} We establish existence and stabilty results for
solitons in noncommutative scalar field theories in even space dimension $2d$.
In particular, for any finite rank spectral projection $P$ of the number
operator ${\mathcal N}$ of the $d$-dimensional harmonic oscillator  and sufficiently large noncommutativity parameter $\theta$
we prove the existence of a rotationally invariant soliton which depends
smoothly on $\theta$ and converges to a multiple of $P$ as $\theta\to\infty$.

In the two-dimensional case we prove that these solitons are stable at large
$\theta$, if $P=P_N$, where $P_N$ projects onto the space spanned by the $N+1$
lowest eigenstates of ${\mathcal N}$, and otherwise they are unstable. We also
discuss the generalisation of 
the stability results to higher dimensions. In particular, we prove
stability of the soliton corresponding to $P=P_0$ for all $\theta$ in its
domain of existence. 

Finally, for arbitrary $d$ and small values of
$\theta$, we prove without assuming rotational invariance that there
do not exist any solitons depending smoothly on $\theta$.

 \vfill

 \newpage
 \pagestyle{plain}

 \section{Introduction}  

Recent progress in string theory has stimulated interest in
solitons in noncommutative field theories
\cite{hoppe,connes,witten}.   
Several authors have found explicit solitons in gauge theories
with and without matter fields \cite{poly,aganagic,bak,gross1,harvey}.   
In \cite{gsm1} solitons in scalar field theories were studied and 
it was shown  that in the case of an infinite
noncommutativity parameter $\theta$, 
where the kinetic term in the action can be neglected, 
large families of solitons exist.
This is in a stark contrast to the commutative case where there are no
solitons \cite{derrick}.
Various aspects of solitons in noncommutative scalar
field theories are discussed in 
\cite{zhou,gorsky,lindstrom,solovyov,bak2,matsuo,gopakumar,araki,anastasia}.
For background and a recent review of some of these results, 
see \cite{komaba}.

The problem we discuss can be formulated either in terms of functions on
${\bf R}^{2d}$, or, by applying a quantization map, in terms of operators
on $L^2({\bf R}^d)$, as explained e.g.\ in \cite{gsm1,komaba}.   In this
paper we do not make use of the former formulation, except for some technical
purposes in the final section. Thus we define {\it solitons} as critical
points of the energy functional
$$
S(\vp)= {\rm Tr}\left(\sum_{k=1}^d[\vp,a_k^*][a_k,\vp] + \theta V(\vp)\right)\;,
$$
where $a_k$ and $a_k^*$ are the standard annihilation and creation operators
of the $d$-dimensional harmonic oscillator, $V$ is a potential, $\theta$ a
positive parameter (called the noncommutativity parameter), and $\vp$ is
a self-adjoint operator on $L^2({\bf R}^d)$. 

In \cite{paper} we established the existence of spherically symmetric solitons
in even dimensional scalar field theories under fairly general conditions
on the potential, provided $\theta$ is sufficiently large and we proved that
no spherically symmetric solutions can exist for small $\theta$.  

Throughout the present paper we assume that {\it $V$ is twice continuously
  differentiable and positive, except for a second order zero at
  $x=0$. Furthermore, we assume that $V'(x)$ is strictly negative for $x<0$
  and has exactly two zeroes at positive values $t$ and $s$ corresponding
 to a local maximum and a local minimum of $V$}, see Fig.\ 1. The techniques
 developed here can be adapted to potentials with more local maxima
 and minima. For the proof of Theorem 5 and for the discussion of stability in
 higher dimensions, we shall assume that $V$ is analytic, although this
 assumption can presumably be relaxed.

Our  results can be divided into two classes, one concerning general solitons
and another concerning solitons
that are diagonal in
the harmonic oscillator basis consisting of the joint eigenfunctions of
$a_k^*a_k$.
In the $d=1$ case the latter solitons correspond to rotationally invariant
functions under the quantization map but in higher dimensions these
solitons correspond to functions that are invariant under rotations in each
of the $d$ quantization planes.  For $d>1$ the rotationally invariant
solitons are those which are functions of the number operator
 ${\mathcal N}$.

In the first category we have the following results for any nonzero
critical point $\vp$ of $S$:

\begin{itemize}
\item   $\vp$ is a positive operator, whose operator norm satisfies
$$
\Vert\vp\Vert\leq s
$$
independently of the value of $\theta$.

\item  $\vp$ is of trace class and ${\rm Tr}\, V'(\vp)=0$.

\item  There exists a nonzero constant $c$ depending only on the potential
  $V$ such that the Hilbert-Schmidt norm of $\vp$, denoted
 $ \Vert\vp\Vert_2$,
  satisfies
$$
\Vert\vp\Vert_2\geq c\theta^{-\frac{d}{2}}\;.
$$
As a corollary  we find that any family  $\vp_\theta$ of
solitons depending smoothly on the noncommutativity parameter $\theta$ 
(in a sense made precise in Section
3)
has a
diverging energy at some strictly positive value of $\theta$.
Hence, such families cannot exist for arbitrarily small values of $\theta$.
This result can be viewed as a noncommutative version of Derrick's theorem
\cite{derrick}.

\end{itemize}

Of results in the second category we mention, in particular, the following.

\begin{itemize}
\item For any finite rank spectral projection $P$ of the number operator
  ${\mathcal N}=\sum_{k=1}^da_k^*a_k$ there exists a maximal smooth family
$$
(\theta_P,\infty )\ni \theta\mapsto\vp_\theta
$$
of solitons 
such that $V''(\vp_\theta)> 0$ and 
$$
\vp_\theta \to sP\quad\mbox{as}\quad \theta\to\infty\;.
$$

\item If $d=1$ and $P$ equals the projection $P_N$ onto the space spanned 
by the $N+1$ lowest eigenstates of ${\mathcal N}$, the solitons $\vp_\theta$ are
  stable for $\theta$ sufficiently large. For all other $P$ the corresponding
  solitons are unstable in their full range of existence. 

\item For $P=P_0$ the corresponding solitons are stable for all $d\geq 1$ in 
 their full range of existence. 

\end{itemize}

This paper is organized as follows.
In a preliminary section we describe the mathematical setting of 
the problem, recall results from
\cite{paper} and prove some technical results on general properties of
solitons. 

In Section 3 we establish the main existence theorem for
solitons. We actually give two proofs, one elementary, generalizing
\cite{paper},
based on an analysis of the difference equation for the eigenvalues of $\vp$
obtained from the Euler-Lagrange equation for the variational problem for $S$,
and another proof based on an application of a fixed point theorem.  While less
elementary, the latter approach has the advantage of giving smoothness of the
solitons as a function of $\theta$. A related existence proof has been
obtained independently in \cite{robert}. 

The results on stability are proven in Section 4, which also contains a
discussion of the extension of our approach to higher dimensions
without giving full details, except in the case $P=P_0$.

Finally, in  Section 5 we prove non-existence of smooth families of solitons
 for small values of $\theta$.  
It should be stressed  that this result only rules out the existence of
smooth families contrary to the nonexistence theorem in \cite{paper} for
rotationally invariant solitons which rules out the existence of any
rotationally invariant solitons for $\theta$ smaller than some positive
$\theta_0$ depending only on $V$ and $d$. It is an interesting unsolved
question whether this stronger result also holds without the assumption of
rotational invariance .

Another interesting unsolved problem concerns existence of general
non-rotation\-ally
invariant solutions, in particular the so called multi-soliton solutions
described in \cite{gsm1}.
The solitons discussed in this paper are special
cases corresponding to overlapping solitons sitting at the origin.  In
\cite{gopakumar} and \cite{lindstrom2} properties of moduli spaces of
multi-solitons are discussed perturbatively in $\theta^{-1}$.  The latter paper
contains a discussion of stability  perturbatively  to first order in $\theta^{-1}$. 
Stability of 
scalar solitons under radial fluctuations is also discussed in \cite{jackson}.
 
\section{General properties of solitons} 

Solitons in a noncommutative $2d$-dimensional scalar field theory with 
a potential $V$  are finite energy solutions to the variational 
equation of the energy functional
\beq{5}
S ( \vp )={\rm Tr}\left(\sum_{k=1}^d [\vp ,a_k^*][a_k,\vp]+\theta V(\vp 
)\right),
\eeq 
where $a_k^*$ and $a_k$ are the usual raising and lowering operators of the
$d$-dimensional simple 
harmonic oscillator and $\vp$ is a self-adjoint operator on
$L^2(\bR^d)$.
We assume that the potential $V$ is at least twice continuously differentiable
with a second order zero at $x=0$
and that $V(x) >0$ if $x\neq 0$. Hence, finiteness of the potential energy
$\theta{\rm Tr}V(\vp)$ requires $\vp$ to belong to the space ${\mathcal H}_2$
of Hilbert-Schmidt operators. Consequently, $S$ is defined and finite on
the space  $ {\mathcal H}_{2,2}$ of self-adjoint Hilbert-Schmidt operators
$\vp$ for which $[a_k,\vp]$ 
is also Hilbert-Schmidt. We note that  ${\mathcal H}_{2,2}$ is a Hilbert space
with norm $\Vert\cdot\Vert_{2,2}$ given by 
\begin{equation}
    \label{H1}
\Vert\vp\Vert^2_{2,2} = \sum_k \Tr ([\vp,a^*_k][a_k ,\vp]) + \Tr \vp^2 =\sum_k
\Vert[a_k,\vp]\Vert^2_2 + \Vert\vp\Vert^2_2\;,
\end{equation}
where $\Vert\cdot\Vert_2$ denotes the Hilbert-Schmidt norm.
It is easy to see that the space ${\mathcal H}_0$ consisting of operators
that are represented by finite matrices (i.e.\ matrices with only finitely many
non-zero entries) in the standard harmonic oscillator basis form a dense
subspace of ${\mathcal H}_{2,2}$.

The variational equation of the functional \rf{5} is 
\beq{6}
2\sum_{k=1}^d [a_k^*,[a_k,\vp ]]=-\theta V'(\vp ).
\eeq
We regard this equation as an equality between  two Hilbert-Schmidt
operators on $L^2 (\bR^d) $. Thus, a solution $\vp$ to Eq.\ \rf{6} 
belongs to
${\mathcal H}_{2,2}$ and has the 
property that the left hand side of Eq.\ \rf{6},
interpreted as a quadratic form on the domain of ${\mathcal N}^\oh$, where ${\mathcal N}$ denotes
the number operator 
$$
{\mathcal N}=\sum_{k=1}^d a_k^*a_k\;,
$$
is Hilbert-Schmidt. We denote the space of such operators by 
${\mathcal D}$. Alternatively, 
 ${\mathcal D}$ is the space of operators $\vp$ in ${\mathcal H}_{2,2}$ such
 that the linear form  
  \begin{equation}
    \label{H2}
    {\mathcal H}_{2,2}   \ni \psi \mapsto  \sum_k \Tr ([a^*_k ,\psi
    ][a_k ,\vp ]) 
  \end{equation}
is continuous in the Hilbert Schmidt norm $\Vert\cdot \Vert_2$.

This operator theoretic formulation of the problem is the most convenient one
for our discussion of the existence and stability results in Sections 3 and
4. For the non-existence results in Section 5 we shall also make
use of
the alternative formulation in terms of ordinary functions and a quantization
map (see e.g.\ \cite{komaba}).
%For the sake of simplicity we shall mostly restrict our attention to the case
%$d=2$ in the remainder of this and the two subsequent sections. The straight
%foreward  generalizations of the results to arbitrary even dimensionality will
%be commented on at appropriate places,
Choosing the harmonic oscillator eigenstates $| n_1 ,\ldots , n_d \kt $,
$n_i=0,1,\ldots $, $a_k^*a_k|n_1,\ldots ,n_d\kt=n_k |n_1,\ldots ,n_d\kt$,
as the basis for the
Hilbert space $L^2(\bR^{d})$, rotationally symmetric functions
correspond, under the standard Weyl quantization, to
diagonal operators whose eigenvalues only depend on
$n_1+\cdots +n_d$.
If we consider a diagonal operator with eigenvalues $\lambda_n$, $n=0,1,2,
\ldots $, Eq.\ \rf{6}
reduces, for $d=1$,  to \cite{gsm1,zhou}
\begin{eqnarray}\label{eqm}
(n+1)\lambda_{n+1}-(2n+1)\lambda_n +n\lambda_{n-1} & = & 
{\theta\over 2} V'(\lambda _n), ~~n \geq 1 \label{em}\\
\lambda_1-\lambda_0 & = & {\theta\over 2} V'(\lambda_0 ).\label{em0}
\end{eqnarray}
Summing the second order finite
difference equation for $\lambda_n$ from $n=0$ to $n=m$ 
yields the first order equation
\beq{7}
\lambda_{m+1}-\lambda_m={\theta\over 2(m+1)}\sum_{n=0}^mV'(\lambda _n),
~~m\geq 0.
\eeq
A necessary condition for the energy to be finite is clearly that
\beq{bc}
\lambda_m\to 0 ~~\mbox{\rm  as}~~ m\to\infty.
\eeq
Actually, this condition implies $\vp\in{\mathcal H}_{2,2}$ by Lemma 1 below.
In \cite{paper} we proved the existence of solutions to Eq.\ \rf{7} satisfying 
the boundary condition \rf{bc} under fairly general conditions on the
potential $V$.  In the next section we generalize that result.

In addition to the conditions on $V$ which have been imposed 
above we assume that $V$ has
only one local minimum in addition to $x=0$. 
Let the other local minimum be at $s>0$.  
Let $r\in (0,s)$ be a point where $V$ has a local maximum and for technical
convenience assume that $V'$ does not vanish except at $0,r$ and $s$.
Then $V'(x)<0$ for $x<0$ or $x\in (r,s)$ and $V'(x)>0$ for $x>s$ or
$x\in (0,r)$ (see Fig. 1).

\begin{figure}[thb]
  \begin{center}
    \psfrag{V'(x)}{$V'(x)$}
   \psfrag{x}{$x$} \psfrag{w}{$w$} \psfrag{t}{$t$}
  \psfrag{r}{$r$}
 \psfrag{s}{$s$}
    \includegraphics[width=10cm]{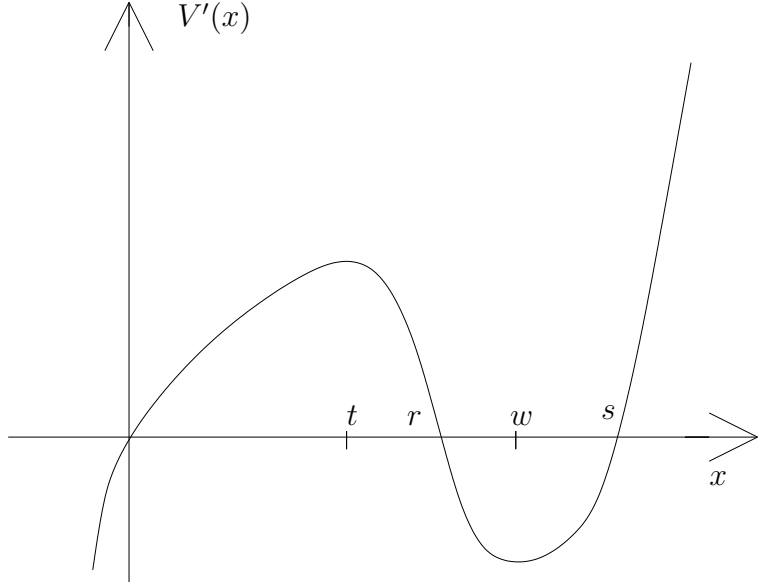}
    \caption{A graph of the derivative of a generic potential $V$
which satisfies our assumptions.}
    \label{fig1}
  \end{center}
\end{figure}

The following result which will be needed in the
next section was proven in  \cite{paper}. We state the result for $d=1$, but
its generalisation to arbitrary $d\geq 1$ is straightforward as explained
in \cite{paper}.

\medskip
\noindent
{\bf Lemma 1.} 
{\it Let $\{\lambda_m\}$ be a sequence of real numbers which 
satisfy Eq.\ \rf{7}.  If $\lambda_n>s$ for some $n$ then $\{\lambda_m\}$ is
increasing for $m\geq n$ and $\lambda_m 
\to\infty$ as $m\to\infty$.
If $\lambda_n\leq 0$ for some $n$ then 
$\{\lambda_m\}$ is decreasing for $m\geq n$ and 
$\lambda_m
\to -\infty$ as $m\to\infty$.

If the sequence $\{\lambda_m\}$ also satisfies 
the boundary condition \rf{bc} and
the $\lambda_m$'s are not all zero then
\begin{itemize}
\item[(i)] $0< \lambda_m < s$, for all $m$.
\item[(ii)] $\lambda_m$ tends monotonically to $0$ for $m$ large enough.
\item[(iii)] $\sum_mV'(\lambda _m)=0$ and  $\sum_m\lambda_m<\infty$.
\end{itemize}}

Dropping the assumption of rotational symmetry we have the following
generalization of (i) and (iii), which, apart from  being of some independent
interest, we will use in Section 5.  The remainder of the present section
is not needed for the existence and stability results 
in the following two sections.

\medskip
\noindent
{\bf Lemma 2.}
{\it  Let $\vp$ be a  nonzero solution to Eq. \rf{6}. Then
\begin{itemize}
\item[(i)]  the operator $\vp$ is positive and its norm
  satisfies the inequality

\beq{a1}
   \Vert\vp \Vert   \leq s .
\eeq

\item[(ii)]  
$\vp$ is of trace class and $\Tr (V'(\vp ))=0$.
\end{itemize}}

\medskip

Before proving the above lemma
 we need the following result, where $\vp_\pm$ denote the
positive and negative parts of a bounded selfadjoint operator $\vp$, defined by

\beq{gggg}
\vp=\vp_+-\vp_-\;,\;\;\vp_+\vp_-=0\;,\;\;\vp_\pm\geq 0\;.
\eeq

\medskip

\noindent
{\bf Lemma 3.} {\it The maps
$$
\vp \mapsto \vp_\pm
$$
are well defined and continuous from ${\mathcal H}_{2,2}$ to itself.}

\medskip

\noindent
{\bf Proof.} 
 Since 
\beq{a2}
\Vert  \vp_\pm \Vert  _2 \leq \Vert  \vp \Vert  _2 ,
\eeq
it suffices to show that, for all $k$, 
\beq{a3}
\Vert  [a_k , \vp_\pm ]\Vert  _2 \leq \mbox{ const~\ } \Vert  [ a_k ,\vp ]\Vert  _2 .
\eeq
We will prove below that this holds with the constant equal to $\sqrt{3}$.
Since ${\mathcal H}_0$ is dense in ${\mathcal H}_{2,2}$ we can assume
$\vp\in{\mathcal H}_0$. It is clear that
the spectral projections of finite rank operators 
corresponding to non-zero eigenvalues 
 belong to  ${\mathcal H}_0$
and the same applies to the spectral projections of
 $\vp_\pm$. In order to estimate the norms
of $\vp_\pm$ it is convenient to write 
\beq{a4}
  \vp_+ = \frac{1}{2\pi i}\int_\gamma \frac{z}{z -\vp }\, dz\;,
\eeq
where $\gamma$ is a  simple closed positively oriented contour in the
complex plane enclosing the positive eigenvalues $\{\lambda_i\}$ of $\vp$ but
 not the non-positive eigenvalues $\{\mu_j\}$. Then 
\begin{equation}
  \label{comm}
  [a_k ,\vp_+ ]=\frac{1}{2\pi i}\int_\gamma \frac{1}{z -\vp } 
[a_k ,\vp] \frac{1}{z -\vp }  z dz \;.
\end{equation}
 Denoting the spectral projection corresponding to $\lambda_i$ by $e_i$ and
 the one of $\mu_j$ by $f_j$, we have 
\beq{a6}
\frac{1}{z-\vp}=\sum_i\frac{1}{z-\lambda_i}e_i + \sum_j\frac{1}{z-\mu_j} f_j\;.
\eeq
Inserting the above identity into Eq.\ \rf{comm} and 
computing residues one obtains
\beq{a7}
  [a_k ,\vp_+ ]= e_+ [a_k ,\vp ]e_+ + \sum_{i,j}
  \frac{\lambda_i}{\lambda_i-\mu_j }(e_i [a_k ,\vp ] f_j + f_j [a_k
  ,\vp ] e_i )\;,
\eeq
where $e_+=\sum_ie_i$ is the support projection of $\vp_+$.
Hence,

\bea
 \Tr ([a_k ,\vp_+ ]^*  [a_k ,\vp_+ ] ) & = & 
                    \Tr ( e_+ [a_k ,\vp ]^* e_+ [a_k ,\vp ]e_+)  \nonumber\\
&& 
\!\!\!\!\!\!\!\!\!\!\!\!\!\!\!\!\!\!\!\!
+\sum_{i,j} \left(\frac{\lambda_i}{\lambda_i-\mu_j }\right)^2
   \Tr (e_i [a_k ,\vp ]^* f_j [a_k ,\vp ] e_i 
                   + f_j [a_k ,\vp ]^* e_i [a_k ,\vp ] f_j) \nonumber   \\  
 &\leq & \Tr ( e_+ [a_k ,\vp ]^* [a_k ,\vp ]e_+ ) \nonumber     \\           
 && +\sum_{i,j}  \Tr (e_i [a_k ,\vp ]^* f_j [a_k ,\vp ]e_i
                    + f_i [a_k ,\vp ]^* e_j [a_k ,\vp ] f_i)\nonumber   \\  
 & \leq & 3\,\Tr ( [a_k ,\vp ]^* [a_k ,\vp ] ),
\eea 
where we used the fact that 
\beq{a8}
0\leq \frac{\lambda_i}{\lambda_i-\mu_j } \leq 1.
\eeq
Clearly, the same estimate applies to  $\Tr ([a_k ,\vp_- ]^*[a_k ,\vp_- ] )$
and the claimed result follows.          

\medskip  

\noindent
{\bf Proof of Lemma 2}.  (i)  We first show that $\vp\geq 0$. Suppose on the
contrary that $\vp_-\neq 0$.  Then, since $V'(-\vp_- ) < 0$, we have for
any integer $n>2$ that 
\begin{equation}
  \label{ineq}
2\sum_{k=1}^d \Tr (\vp_-^n [a_k^* , [\vp , a_k]]) =
\theta \Tr ( \vp_- ^n V'(\vp ))<0\;.  
\end{equation}
But, using the cyclicity of the trace,
\bea
 \Tr (\vp_-^n [a_k^* , [\vp , a_k]]) &
   = & \Tr ( \vp_-^n [a_k^* , [\vp_+ , a_k]]) - \Tr (\vp_-^n [a_k^* , [\vp_- ,
   a_k]] )\nonumber\\ 
 & =& \Tr ([a_k^* ,\vp_-^n ][\vp_- ,a_k]) - 
\Tr ([a_k^* ,\vp_-^n ][\vp_+ ,a_k])\nonumber\\
 &  =  & \sum_{p+q=n-1} \Tr (\vp_-^p [a_k^* ,\vp_- ] \vp_-^q [\vp_- ,a_k])
\nonumber\\ 
&&  -\sum_{p+q=n-1}\Tr (\vp_-^p [a_k^* ,\vp_- ]\vp_-^q [\vp_+ ,a_k]) 
\nonumber\\
 &  = & \sum_{p+q=n-1} \Tr (\vp_-^\frac{p}{2} [a_k^* ,\vp_- ] \vp_-^q
  [\vp_- ,a_k] \vp_-^\frac{p}{2})   \nonumber \\
&&\!\!\!\!\!\!\!\!\!\!
\!\!\!\!\!\!\!\!\!\!
\!\!\!\!\!\!
  +   \Tr (\vp_+^\frac{1}{2} [\vp_- ,a_k]\vp_-^{n-2}[a_k^* ,\vp_-]
  \vp_+^\frac{1}{2}) + \Tr (\vp_+^\frac{1}{2}  [a_k^*
  ,\vp_-]\vp_-^{n-2}  [\vp_- ,a_k]  \vp_+^\frac{1}{2} )
\nonumber\\ 
& \geq & 0\;,
\eea
which contradicts the inequality \rf{ineq}.  

To prove the inequality in (i) we note that the equation of motion \rf{6}
implies  that
\begin{eqnarray}
\Vert  \vp \Vert  ^{-n} \theta \Tr (\vp^n  V'(\vp )) &=& 2\sum_k\Tr (\Vert  \vp \Vert  ^{-n}
\vp^n [a_k^* ,[\vp ,a_k ]])\nonumber\\
&=&-2\sum_k\Vert  \vp \Vert  ^{-n} \Tr [a_k^* ,\vp_n ][\vp ,a_k ] < 0.
\end{eqnarray}
We also have
\beq{a9}
 \lim_{n\rightarrow \infty} \Vert  \vp \Vert  ^{-n} \theta \Tr (\vp^n
 V'(\vp ))= \theta  V'(\Vert   \vp \Vert  ) \Tr e\;,
\eeq
where $e$ is the spectral projection of the operator $\vp$
corresponding to the eigenvalue $\Vert  \vp \Vert  $. In particular,
\beq{a10}
  \theta  V'(\Vert   \vp \Vert  ) \Tr e \leq 0\;, 
\eeq
which implies the desired inequality by the assumed form of the potential $V$.

(ii)  Let $P_m$, $m=0,1,2,\ldots$, denote the orthogonal projection onto the
eigenspace of the number operator ${\mathcal N}$ corresponding to eigenvalue $m$, and set
\beq{a11}
\lambda_m =\Tr (P_m \vp )\;.
\eeq
Then the equation of motion \rf{6}  gives  
\beq{a12}
 \oh\theta \: \Tr (P_m V'(\vp ) ) =
 (m+1)\lambda_{m+1}-(2m+d) \lambda_m + (m+d-1)\lambda_{m-1}.
\eeq
Summing this identity over $m\leq n$ we get (as in the
spherically symmetric case)
\beq{a13}
(n+1)  \lambda_{n+1}- (n+d)\lambda_n  =\theta  \sum_{i\leq n}
  \Tr (P_i V'(\vp ) ) ,
\eeq
and, finally, summing over $n\leq p$,
\begin{equation}
  \label{recur}
 \lambda_{p+1}-\lambda_0 
     = \theta \sum_{n\leq p} \frac{1}{(n+1)}\left( (d-1)\lambda_n
                       + \sum_{i\leq n}   \Tr (P_i V'(\vp ) )\right) .
\end{equation}
Besides this equation we shall also make use of the fact that 
\begin{equation}
  \label{asym}
V' (\vp) =a\vp + O(\vp^2)
\end{equation}
for some positive constant $a$ as a consequence of the assumptions made on
$V$. Since $\vp$ is Hilbert-Schmidt it follows from this that $V'(\vp)$ is of
trace class if and only if $\vp$ is of trace class. We first prove that this
is the case if (and only if) $\lim_{m\rightarrow \infty}\lambda_m =0$
and in this case, $\Tr V'(\vp )=0$.
In fact, by \rf{asym},
\beq{part} 
 \sum_{i\leq n} \Tr P_i (V'(\vp ))
= \sum_{i\leq n}(a \lambda_i  + c_i)\;,
\eeq
where $\sum_i c_i$ is absolutely convergent while all the terms in 
$
\sum_{i\leq n} \lambda_i
$ 
are positive, since $\vp$ is a positive operator by (i).
It follows that the sum $ \sum_{i\leq n} \Tr P_iV'(\vp )$ has a limit
$L$, finite or $+\infty$, as $n\rightarrow\infty$. 
On the other hand, it follows from our assumptions 
that the right hand side of Eq.\ \rf{recur} 
converges as $p\rightarrow\infty$ and consequently, 
since the $\lambda_m$'s are
nonnegative, $L$ must be zero. Hence, Eq.\ \rf{part} implies that 
$\sum_i\lambda_i$ converges, i.e., $\vp $ is of trace
class, and the trace $L$ of $V'(\vp )$ is zero as claimed.

It remains to show that $\lambda_m\to 0$ as $m\to\infty$. Assume this is not
the case. Then $\sum_i\lambda_i=+\infty$ and therefore, by Eq.\ 
\rf{part}, we have
\beq{uuu}
\sum_{i\leq m}\Tr (P_iV'(\vp))>1
\eeq
for $m$ large enough.
Thus, by Eq.\ \rf{recur},
\beq{iop}
 \lambda_{p} \geq \theta \sum_{n\leq p-1} \frac{1}{n+1}\sum_{i\leq n}
\Tr (P_iV'(\vp) ) \geq \mbox{const~}\ln p\;,
\eeq
for $p$ large enough. Repeating the argument with the inductive assumption
$\lambda_p\geq \mbox{const~}p^l$, for sufficienly large $p$, where 
$l$ is a nonegative integer, leads to
$\lambda_p\geq \mbox{const~}p^{l+1}$ for $p$ sufficeiently large. 
Hence, $\lambda_m$ increases faster
than any power of $m$, if it does not tend to zero. But this is not possible 
since, by the Cauchy-Schwarz inequality,
\beq{pop}
 \lambda_m^2 =(\Tr P_m \vp )^2 \leq \Tr ( P_m \vp^2 ) \Tr P_m 
\leq\mbox{const~}m^{d-1} \Tr ( P_m \vp^2 )
\eeq
and hence,
\beq{uyt}
\sum_m \frac{ \lambda_m^2}{m^{d-1} }\leq \sum_m \Tr ( P_m \vp^2
)=\Vert  \vp\Vert  _2^2 < \infty\;.
\eeq
This finishes the proof of Lemma 2.

\medskip

\section{Existence}

We now proceed to discuss the existence of rotationally invariant solutions to
Eq.\ \rf{6}. 
Let $t$ be the location of the maximum of 
$V'$ in the interval $[0,s]$ and let $w$ be the location of the 
minimum of $V'$ in the same interval (see Fig.\ 1).  As above we denote by
$P_0, P_1,\dots$ the orthogonal projections onto the eigenspaces of the number
operator of the $d$-dimensional harmonic oscillator. The purpose of this
section is to prove the following theorem.

\medskip
\noindent
{\bf Theorem 1.} {\it For any projection $P$ on $L^2({\bf R}^d)$, which is the sum
of a finite number of the projections $P_n$, there is a unique maximal family
$\vp_\theta,\; \theta>\theta_P$, of rotationally invariant solutions of
Eq.\ \rf{6}, which depends smoothly on $\theta$, i.e., is continuously
differentiable with respect to the norm $\Vert\cdot\Vert_{2,2}$, and fulfills 
\beq{yyt}
V''(\vp_\theta)> 0\;,
\eeq
as well as
\beq{lim}
\vp_\theta\to s\,P
\eeq
in Hilbert-Schmidt norm as $\theta\to\infty$.}

\medskip
\noindent
{\bf Proof.} We shall give two proofs of existence of solutions for
sufficiently large $\theta$. 
The first proof is an extension of the proof given in  \cite{paper} for
$P=P_0$. For simplicity we restrict to $d=1$ and to $P=P_0+\cdots P_N$, the
adaptation of the arguments to arbitrary  $d\geq 1$ being explained in
\cite{paper}.   

First, assume $\theta$ is so large 
that

\beq{13}
{\theta\over 2(N+1)}| V'(w)|\geq w.
\eeq
In this case we claim there is a unique 
$\underline{\lambda}\in [w,s)$ such that if 
we set $\lambda_0=\underline{\lambda}$ and define $\lambda_i$ for $i>0$ 
by the recursion \rf{7} then
\beq{order}
\lambda_0>\lambda_1>\ldots \lambda_N\geq w
\eeq
 and $\lambda_{N+1}=0$.
In order to prove the claim we begin by choosing $\lambda_0$ close to 
but smaller than $s$ so that \rf{order} holds, which clearly is possible.  Then 
$\lambda_N>\lambda_{N+1}$ by \rf{7}, and if $\lambda_{N+1}=0$ 
we are done.  Note that all the $\lambda_i$'s 
depend continuously on $\lambda_0$ and
$\lambda_{N+1}\to s$ as $\lambda_0\to s$.  
If $\lambda_{N+1}<0$ we increase $\lambda_0$ until $\lambda_{N+1}=0$  and the 
inequalities \rf{order} still hold because $\lambda_1,\ldots \lambda_N$ all 
increase with $\lambda_0$.  If $\lambda_{N+1}>0$ we decrease $\lambda_0$ until
$\lambda_{N+1}=0$ and \rf{order} still holds due to the inequality \rf{13}.
This proves the existence of $\underline{\lambda}$.

Next take $\theta$ still larger, if necessary, so that
\beq{ii}
V'(t)\geq (N+1)|V'(\underline{\lambda})|.
\eeq
This is clearly 
possible because $\underline{\lambda}\to s$ as $\theta\to\infty$.
We now claim there exists $\bar{\lambda}\in (\underline{\lambda},s)$ such 
that if we take $\lambda_0=\bar{\lambda}$ then \rf{order} holds and
$\lambda_{N+1}=\lambda_{N+2}$, i.e.
\beq{q1}
0={\theta\over\ 2 (N+2)}\sum_{i=0}^{N+1}V'(\lambda_i).
\eeq 
In order to verify the existence of $\bar{\lambda}$ we note that, as a
consequence of \rf{7}, for 
$\lambda_0$ greater than but close to $\underline{\lambda}$ we have 
$\lambda_{N+1}$ is greater than but close to $0$, and $\lambda_{N+1}$ increases with $\lambda_0$.  
Hence, in view of \rf{ii} and the fact that $\lambda_1,\dots,\lambda_N$ are
also increasing functions of $\lambda_0$, there is a
$\lambda_0\equiv\overline\lambda\in (\underline{\lambda},s)$ such that
\beq{q2}
V'(\lambda_{N+1})=-\sum_{i=0}^NV'(\lambda_i)
\eeq
which establishes the claim.  We note that for $\lambda_0=\overline{\lambda}$
we have $\lambda_{N+1}\in (0,t)$.

  If a sequence $\{\lambda_i\}$
obeys the recursion \rf{7} and has the property
$\lambda_0>\lambda_1>\ldots > \lambda_p$, but $\lambda_{p+1}\geq \lambda_p$,
we say that the sequence {\em turns at} $p$.  
We note that in this case $\lambda_p>0$ by Lemma 1
and if $\lambda_{p+1}=\lambda_p$ then $\lambda_{p+2}>\lambda_{p+1}$ by \rf{7}.

Define the set
\beq{18}
A=\{\lambda_0\in [\underline{\lambda},\bar\lambda ]:
\{\lambda_i\} ~~\mbox{\rm turns at some}~~ p\}.
\eeq
By construction $\underline{\lambda}\notin A$ and $\bar\lambda\in A$.
Put $\Lambda_0=\inf A$.  Since each $\lambda_i$ depends continuously
on the initial value $\lambda_0$ it follows that $\Lambda_0\notin A$.

Now consider the sequence defined by $\lambda_0=\Lambda_0$ and
Eq.\ \rf{7}.  Since this
sequence does not turn it is monotonically decreasing.  In order to
show that this sequence provides 
a solution to our problem it therefore suffices to 
show that $\lambda_i\to 0$ as $i\to\infty$.  
Suppose $\lambda_i$ becomes negative for some $i$.
Then Lemma 1 implies that $\lambda_i\to -\infty$.
By the continuity of $\lambda_i$ as a function of $\lambda_0$ it 
follows that for $\lambda_0$ sufficiently close to $\Lambda_0$
the sequence $\lambda_i$ 
tends monotonically to $-\infty$  but this contradicts the definition 
of $\Lambda_0$.  
We conclude that the limit $\lim_{i\to\infty}\lambda_i=a\geq 0$ exists
and by \rf{7} we have
\beq{19}
V'(a)={2\over \theta}\lim_{i\to\infty}(\lambda_{i+1}-\lambda_i)=0.
\eeq
Hence, $a=0$ since $\lambda_i<r$ for $i>N$. 
This completes the proof of the existence of rotationally invariant solutions
$\vp_\theta$ for large enough $\theta$ and it follows easily from the
construction that $\vp_{\theta}\to sP$ in operator norm as $\theta\to\infty$.

It is worth while noting that the proof given here shows that the sequence of
eigenvalues $\{\lambda_i\}$ of $\vp_\theta$ is strictly decreasing for $\theta$
large enough. This is special for the choice of projection $P$ made above.
The same technique can be applied to
demonstrate existence of solutions converging to any projection of the type
stated in the theorem, but since this result as well as the  claim of
differentiability are obtained in a more uniform manner by the second method of
proof, we shall not discuss that approach in more detail here.
Also, the above proof can easily be generalized to establish the existence of
solutions which converge to finite rank operators of the form $tP+sP'$,
$PP'=0$, as $\theta\to\infty$.

The second proof of existence is by use of a fixed point theorem. Let us
first note that the operator $\Delta$, defined by
\beq{delta}
\Delta\vp=\sum_{k=1}^d[a_k^*,[a_k,\vp]]\;,
\eeq
is self-adjoint and positive on ${\mathcal H}_2$ with domain
${\mathcal D}$. Indeed, as explained in Section 5, it is unitarily equivalent
to the standard Laplace operator on $L^2({\bf R}^{2d})$ via a quantization map
$\pi_W :L^2({\bf R}^{2d})\to {\mathcal H}_2$, which justifies the notation $\Delta$ for
this operator in the remainder of this proof. Given a  bounded self-adjoint
operator $B$ on $L^2({\bf R}^{d})$, it defines by left multiplication a bounded
self-adjoint operator on ${\mathcal H}_2$, which we shall also denote by $B$. By the
Kato-Rellich  theorem  $\Delta + B$ is self-adjoint with domain
${\mathcal D}$. Assuming $B\geq c>0$ we have $\Delta + B\geq c$ and hence $\Delta + B$ maps
${\mathcal D}$ bijectively onto ${\mathcal H}_2$ with bounded inverse    
\beq{z1}
(\Delta + B)^{-1}\leq c^{-1}\;.
\eeq
The same statement holds if $B$ is of the form 
\beq{dia}
B=\sum_{n=0}^\infty b_n P_n
\eeq
and we restrict $\Delta +B$ to ${\mathcal D'}={\mathcal D}\cap {\mathcal H}_2'$, where
${\mathcal H}_2'$ is the Hilbert subspace of ${\mathcal H}_2$
consisting of diagonal operators of the form \rf{dia}. This follows by using
that ${\mathcal H}_2'$  corresponds under the quantization map $\pi_W$ to rotation
invariant functions in $L^2({\bf R}^{2d})$ on which the Laplace operator is
known to be self-adjoint. Alternatively, one can use the explicit form 
\beq{z2}
\Delta\vp=-\sum_{n=0}^\infty\{(n+d)\lambda_{n+1}-(2n+d)\lambda_n
  +n\lambda_{n-1}\} P_n \;,
\eeq
where $\vp=\sum_{n=0}^\infty \lambda_{n} P_n$, and the domain ${\mathcal D'}$
consists of those $\vp$ which fulfill 
\beq{z3}
\sum_{n=0}^\infty |(n+d)\lambda_{n+1}-(2n+d)\lambda_n
  +n\lambda_{n-1}|^2 <\infty\,.
\eeq
Since $\Delta + B$ is a closed symmetric operator it suffices to verify
that the orthogonal complement to its range is $\{0\}$. But it is easily seen
that $\vp$ belongs to this orthogonal complement if and only if
\beq{z4}
(n+d)\lambda_{n+1}-(2n+d)\lambda_n +n\lambda_{n-1} =
b_n\lambda_n,
\eeq
for $n\geq 0$.
The proof of Lemma 1 shows that any non-trivial solution $\{\lambda_n\}$ of
this recursion relation  diverges to $\pm\infty$, since $b_n\geq c >0$. Hence
$\vp=0$ if $\vp\in {\mathcal H}_2'$, as desired.

As a consequence, we note that for $\rho\geq 0$ and $B$ and $c$ as above,
the operator $\rho\Delta + B$ has a bounded inverse on ${\mathcal H}_2'$ fulfilling 
\beq{bound}
(\rho\Delta + B)^{-1}\leq c^{-1}\;,
\eeq
the case $\rho=0$ being obvious.

In view of these preparatory remarks, we rewrite Eq.\ \rf{6} as
\beq{fix1}
\rho\Delta\vp + V'(\vp) =0\;,
\eeq  
where $\rho=2\theta^{-1}$. Then $\psi_0= sP$ is a solution for $\rho=0$.
Since $\psi_0\in {\mathcal H}_2'$ and 
\beq{z7}
V''(\psi_0)\geq \mbox{min}\{V''(0),V''(s)\}\equiv c_0>0\;,
\eeq
by assumption, we can, for $\rho\geq 0$, further rewrite the equation in the
form 
\beq{fix2}
\vp= (\rho\Delta +V''(\psi_0))^{-1}\{V''(\psi_0)\psi_0 + 
V'(\psi_0)-V'(\vp)-V''(\psi_0)(\psi_0-\vp)\}
\equiv T_\rho(\vp)\;.
\eeq
Since $V$ is $C^2$ by assumption we have 
\beq{z8}
\Vert V'(\vp)-V'(\psi_0)-V''(\psi_0)(\vp-\psi_0)\Vert_2
= o(\Vert\vp-\psi_0\Vert_2  )\,,
\eeq
and also
\beq{z9}
\Vert (\rho\Delta +V''(\psi_0))^{-1}V''(\psi_0)\psi_0
-\psi_0\Vert_2 = \rho\Vert(\rho\Delta
+V''(\psi_0))^{-1}\Delta\psi_0\Vert_2\leq c_1\rho\;, 
\eeq
where $c_1= c_0^{-1}\Vert\Delta\psi_0\Vert_2$.

For $\vp$ in the ball
\beq{z11}
B_\varepsilon(\psi_0)=\{\vp\in
{\mathcal H}_2' : \Vert\vp-\psi_0\Vert_2\leq\varepsilon\}\;,
\eeq
we then  have 
\beq{z12}
\Vert T_\rho(\vp)-\psi_0\Vert_2\leq c_1\rho + o(1) \Vert\vp-\psi_0\Vert_2\;,
\eeq
and hence, $ T_\rho(\vp)\in B_\varepsilon(\psi_0)$ if $\rho$ and
$\varepsilon$ are sufficiently small.
Similarly, one sees that 
\beq{z14}
\Vert T_\rho(\vp)- T_\rho(\psi)\Vert_2\leq
o(1) \,\Vert\vp-\psi_0\Vert_2\;, 
\eeq
so $T_\rho$ is a contraction on  $B_\varepsilon(\psi_0)$, if  $\rho$
and $\varepsilon$ are sufficiently small. 
Fixing $\varepsilon$ accordingly, Banach's fixed point theorem implies the
existence of a unique solution $\psi_\rho$ of Eq.\ \rf{fix1} in
$B_\varepsilon(\psi_0)$ for $0\leq\rho\leq\delta$ and $\delta$ small
enough.

For  $0\leq\rho,\;\rho_0\leq\delta$, we have 
\beq{fix3}
\psi_\rho-\psi_{\rho_0} =  (\rho\Delta +V''(\psi_0))^{-1}\{(\rho_0-\rho)\Delta
\psi_{\rho_0} + V'(\psi_{\rho_0})-V'(\psi_\rho
)-V''(\psi_0)(\psi_\rho-\psi_{\rho_0} )\} 
\eeq
from which we get
\beq{z55}
\Vert \psi_\rho-\psi_{\rho_0}\Vert_2\leq c_2|\rho-\rho_0| + o( \Vert
\psi_\rho-\psi_{\rho_0}\Vert_2)\;,
\eeq
where the constant $c_2$ depends only on $\rho_0$, and we have assumed
$\varepsilon$ is small enough such that $V''(\psi_\rho)>0$. This inequality
implies that $\psi_\rho$ is a Lipschitz continuous function of $\rho$ if
$\varepsilon$ is small enough. In turn, Eq.\ \rf{fix3} implies that $\psi_\rho$
is differentiable in the $\Vert\cdot\Vert_2$-norm with 
\beq{der}
\frac{d\psi_\rho}{d\rho}=(\rho\Delta +V''(\psi_0))^{-1}\Delta\psi_{\rho}\;.
\eeq
By standard arguments, the family $\psi_\rho,\;0\leq\rho<\delta$
extends to a maximal family, differentiable in the $\Vert\cdot\Vert_2$-norm, and
such that $V''(\psi_\rho)>0$.

It remains to establish the stronger claim of smoothness in the norm
$\Vert\cdot\Vert_{2,2}$ for $\rho > 0$. In order to obtain this,
it is  sufficient to verify that the bijective operator $(\rho\Delta
+V''(\vp))^{-1}$ from ${\mathcal H}_2$ onto ${\mathcal D}'$ is bounded, 
when is ${\mathcal
  D}'$ equipped with the $\Vert\cdot\Vert_{2,2}$-norm, for $\rho>0$ and
$V''(\vp)>0$.  It is straightforward to verify that under these conditions
$(\rho\Delta +V''(\vp))^{-1}$ is bounded (and $\rho\Delta +V''(\vp)$
as well, in fact), when  ${\mathcal D}'$ is equipped with the norm
\beq{z99}
\Vert\vp\Vert_{4,2} =
\left(\Vert\Delta\vp\Vert_2^2+\Vert\vp\Vert_2^2\right)^{\oh}\;, 
\eeq
which is easily seen to be stronger than
$\Vert\cdot\Vert_{2,2}$. In addition, simple estimates show that the
derivative given by Eq.\ \rf{der} is continuous in this norm.

 This
completes the proof of the theorem
with $\vp_\theta=\psi_\rho$ for $\rho=2\theta^{-1}$.

We remark that the above argument can easily be generalized to prove the
existence of solutions to Eq.\ \rf{6} which converge to $sP$, where $P$ is
a projector onto space spanned by a finite number of the
joint eigenfunctions of the number operators $a_k^*a_k$.  As remarked above,
these solutions are not rotationally invariant but only invariant under
rotations in the $d$ two-dimensional quantization planes.

\section{Stability}

In this section we study the stability of solutions to Eq.\ 
\rf{6} in the case $d=1$. Extension to $d>1$ is briefly discussed at the end
of the section. 

A solution
$\vp$ is defined to be stable if the second functional derivative of the
action $S$  at $\vp$ is a  positive semidefinite quadratic form at $\vp$, i.e.,
\beq{x1}
\Sigma(\omega)\equiv \frac{1}{2}\left.
{d^2\over d\epsilon^2}S(\vp +\epsilon\omega )\right|_{\epsilon=0}\geq 0\;.
\eeq
The natural domain of definition of the quadratic form
$\Sigma$ depends generally both on the
potential $V$ and on $\vp$. Under the previously stated assumptions on $V$
the domain
contains at least the space ${\mathcal H}_0$ for the rotationally symmetric
solutions that we consider here.
 If
$\Sigma$ is continuous with respect to the norm 
$\Vert  \cdot\Vert  _{2,2}$ it is
sufficient to show stability for perturbations $\omega$ in ${\mathcal
  {\mathcal H}}^0$. Since the kinetic term in $S(\vp)$ is quadratic, continuity of
$\Sigma$ means that the second functional derivative of $V$ is a continuous
quadratic form with respect to the Hilbert-Schmidt norm. This continuity is 
easy to check, using the analytic functional calculus, if $V$ is 
analytic in a neighborhood of the interval
$[0,s]$ which we will assume to be the case from now on.  For this reason 
we restrict attention below to $\omega\in{\mathcal
  {\mathcal H}}^0$.   Our results about stability can be summarized in the following
three theorems. 

\medskip
\noindent
{\bf Theorem 2.}
{\it Let $\vp$ be a rotationally invariant, finite energy 
solution to \rf{6} with a nondegenerate spectrum 
and let $\lambda_0,\lambda_1,\ldots $
denote the eigenvalues of $\vp$ in the harmonic oscillator basis.  Then
$\vp$ is unstable unless $\{\lambda_n\}$ is a decreasing sequence.}

\medskip  

This theorem implies that only the solutions corresponding to
$P=P_0+\dots+P_N$ in
Theorem 1 can possibly be stable.  By abuse of notation we denote this solution
by $\vp_N$, for a fixed value on $\theta$, in the remainder of this section.

\medskip

\noindent
{\bf Theorem 3.} {\it The solution $\vp_0$ of Eq.\ \rf{6} constructed in the
  previous section is stable for all values of $\theta$ in the maximal range.} 

\medskip

\noindent
{\bf Theorem 4.}  {\it For any $N\geq 0$ the solution $\vp_N$ constructed in
  the previous section is stable for $\theta$ sufficiently large.}

\medskip

We note that Theorem 3 implies Theorem 4 in the case $N=0$. We choose to
state and prove Theorem 3 separately because it is stronger than Theorem 4
for $N=0$ and the proof is simpler.  In the proof of Theorem 4 we have to
rely on asymptotic expansions of the eigenvalues for large $\theta$ which
are not needed in the proof of Theorem 3.  We remark further that solutions
with eigenvalues $\lambda_n$ some of which lie in the region where $V''<0$
are in general unstable but one can construct examples of stable
solutions with eigenvalues in the region where $V''<0$.  

Before proving the theorems we do some groundwork and establish notation.
Let 
\bea
K(\vp)& = & \Tr [\vp,a^*][a,\vp]\nonumber\\
         & = & \sum_{n,m=0}^\infty |\br n|[a,\vp]|m\kt |^2
\eea
denote the kinetic energy functional.  Let $\vp$ be a rotationally invariant
solution of Eq.\ \rf{6} with a nondegenerate spectrum.
Then we can write 
\beq{x5}
\vp +\epsilon\omega =U_\epsilon^*\vp_\epsilon U_\epsilon
\eeq
where $U_\epsilon$ is unitary and $\vp_\epsilon$ is diagonal in the harmonic
oscillator basis.  It follows that
\beq{x6}
\left.
{d^2\over d\epsilon^2}S(\vp +\epsilon\omega )\right|_{\epsilon=0}
=2K(\omega)+\theta\left. {d^2\over d\epsilon^2} 
\Tr V(\vp_\epsilon)\right|_{\epsilon=0}.
\eeq 
Notice that the assumption $\omega\in{\mathcal H}_0$
implies that only finitely many of the eigenvalues and eigenvectors of $\vp$
are perturbed, and we can apply standard non-degenerate perturbation theory.
Let $\lambda_n(\ep)$ denote the eigenvalue of $\vp_\ep$ which converges to
$\lambda_n$ as $\ep\to 0$.  Then $\lambda_n(\ep )$ is real analytic in $\ep$,
and 
\beq{x7}
\left. {d^2\over d\epsilon^2}
\Tr V(\vp_\epsilon)\right|_{\epsilon=0}=\sum_{n=0}^\infty
\left(\lambda_n''(0)V'(\lambda_n)+(\lambda_n'(0))^2V''(\lambda_n)\right).
\eeq
From standard perturbation theory we know that
\beq{x8}
\lambda_n'(0)=\br n|\omega |n\kt
\eeq
and
\beq{x9}
\lambda_n''(0)=2\sum_{m\neq n} {|\br n|\omega |m\kt|^2\over
\lambda_n-\lambda_m}.
\eeq
The condition for stability can therefore be written as
\bea
\Sigma (\omega) & = & K(\omega)+\theta\sum_{m\neq n} {|\br n|\omega
|m\kt|^2\over
\lambda_n-\lambda_m}V'(\lambda_n) + 
\frac{\theta}{2}\sum_{n=0}^\infty |\br n|\omega
|n\kt |^2V''(\lambda_n)\nonumber\\
& = & K(\omega)+\theta\sum_{m<n} {|\br n|\omega
|m\kt|}^2\frac{V'(\lambda_n)-V'(\lambda_m)}{\lambda_n-\lambda_m} + 
\frac{\theta}{2}\sum_{n=0}^\infty |\br n|\omega
|n\kt |^2V''(\lambda_n)\nonumber \\
       & \geq & 0.\label{x10}
\eea
  We remark that the
last term in $\Sigma$ is nonnegative if $V''(\lambda_n)\geq
0$ for all $n$.  The kinetic energy term can be written
\beq{x11}
\sum_{n,m=0}^\infty |\sqrt{n+1}\,\br n+1|\omega |m\kt-\sqrt{m}\,\br n|\omega
|m-1\kt|^2\;,
\eeq 
where $\sqrt{m}\,\br n|\omega|m-1\kt$ is set to zero for $m=0$,
and we see that the kinetic energy couples the matrix elements of
$\omega$ to their nearest neighbours along diagonals with $n-m$ fixed.
On the other hand, the potential part of $\Sigma$  does not
couple different matrix elements of $\omega$.  Note that $\br n|\omega |m\kt
=\br m|\omega | n\kt ^*$ since $\omega$ is self-adjoint but otherwise 
the matrix elements of $\omega$ can be chosen arbitrarily.

\medskip

\noindent
{\bf Proof of Theorem 2.}  We will show that there exists a perturbation
$\omega$ such that $\Sigma (\omega)<0$ unless the $\lambda_n$'s are
decreasing.  We take $\omega$ such that $\br n |\omega | m\kt =0$ for
$|n-m|\neq 1$.  Then we can write
\bea
\Sigma (\omega ) & = & \sum_{n=0}^\infty \left(|\sqrt{n+1}\,\br n+1 |\omega
|n\kt - \sqrt{n} \,\br n|\omega |n-1\kt |^2 \right.\nonumber \\ 
 & & \left.+
|\sqrt{n+1}\,\br n+1 |\omega
|n+2 \kt - \sqrt{n+2} \,\br n|\omega |n+1\kt |^2\right)\nonumber\\
 & & +\theta\sum_{n=0}^\infty {|\br n|\omega |n+1\kt |^2\over
\lambda_n-\lambda_{n+1}} \left(V'(\lambda_n)-V'(\lambda_{n+1})\right).
\label{x12}
\eea
The above expression is quadratic in the variables
\beq{x13}
\alpha_n=\br n+1|\omega |n\kt ,
\eeq
$n=0,1,2,\ldots $.  Assuming without loss of generality that the
$\alpha_n$'s are real we have
\beq{x14}
\Sigma (\omega )=2\sum_{n,m}q_{nm}\alpha_n\alpha_m\;,
\eeq
where the symmetric matrix $q_{nm}$ has only nonvanishing matrix elements on
the diagonal and next to the diagonal which are given by
\bea
q_{nn} & = & 2(n+1)+\gamma_n\\
q_{nn+1} & = & -\sqrt{n+1}\,\sqrt{n+2}\\
q_{nn-1} & = & - \sqrt{n}\,\sqrt{n+1},
\eea
where
\beq{x16}
\gamma_n={\theta\over 2}
{V'(\lambda_{n+1})-V'(\lambda_n)\over \lambda_{n+1}-\lambda_n}.
\eeq
We need to show that $q_{nm}$ is a positive semidefinite matrix.
This is most easily done by diagonalising $q_{nm}$, using elementary row and
column operations, and verifying that the diagonal entries $C_0, C_1,\dots$ in
the resulting diagonal matrix $C$ are non-negative. In the first
step we divide the first row by $q_{00}$, multiply it by $-q_{10}$ and add
the resulting row to the second row.  Then we see that the first two
diagonal entries of $C$ are
\bea
C_0 & = & q_{00}\\
C_1 & = & q_{11}-{q_{10}^2\over q_{00}}.
\eea
Inductively we find  
\beq{x17}
C_{k} = q_{kk}-{q_{kk-1}^2\over C_{k-1}}.
\eeq
We can evaluate $C_0$ and $C_1$ directly using the equation of
motion \rf{eqm} and find
\bea 
C_0 & = & 2\, {\lambda_2-\lambda_1\over \lambda_1-\lambda_0},\\
C_1 & = & 3\, {\lambda_3-\lambda_2\over \lambda_2-\lambda_1}.
\eea
Now it is straightforward to prove from Eq.\ \rf{x17} by induction that
\beq{x18}
C_k = (k+2)\, {\lambda_{k+2}-\lambda_{k+1}\over \lambda_{k+1}-\lambda_k}
\eeq
and we conclude that $C_k> 0$ for all $k$ if and only if the
sequence $\{\lambda_n\}$ is monotone.  Obviously, the sequence cannot be
increasing since $\lambda_n>0$ for all $n$ and $\lambda_n\to 0$ as 
$n\to\infty$.

\medskip

\noindent
{\bf Proof of Theorem 3.}  Let $\lambda_n$ be the eigenvalue of $\vp_0$ 
corresponding to the eigenvector $|n\kt$, $n=0,1,2,\ldots$.
Since $V''(\lambda_n)\geq 0$ for all $n$, by hypothesis, 
and the kinetic energy only couples the matrix
elements of $\omega$ along diagonals it is sufficient and also necessary, 
in view of Eq.\ \rf{x10}, to prove that
\bea
\Sigma_k(\omega) & \equiv & \sum_{n-m=k}\left( |\br n|[a,\omega ]| m\kt |^2
+|\br m|[a,\omega ]| n\kt |^2  +  \theta
|\br n| \omega |m\kt
|^2\frac{V'(\lambda_n)-V'(\lambda_m)}{\lambda_n-\lambda_m}\right)\nonumber\\ 
  & \geq & 0
\eea
for $k\geq 1$.  For each fixed $k$ the argument is quite similar to the proof
of the previous theorem.  We put $\alpha_n=\br n+k |\omega |n \kt$ which can
be assumed to be real for the purpose of proving positivity.  We see that
$\Sigma_k(\omega) $ is a quadratic form $2Q_k$ in the variables 
$\alpha_n$. As in the previous proof the matrix representing $Q_k$ has only
nonvanishing matrix elements  on the diagonal and next to it, and they are
given by
\bea
q_{nn} & = & 2n +1+k+\gamma_n\\
q_{nn-1} & = & -\sqrt{n(n+k)}\\
q_{nn+1} & = & -\sqrt{(n+1)(n+1+k)}
\eea
and
\beq{x20}
\gamma_n={\theta\over 2} {V'(\lambda_n)-V'(\lambda_{n+k})\over \lambda_n-\lambda_{n+k}}.
\eeq
The positivity of this form is equivalent to the positivity of the numbers
$C_n$ defined inductively by
\beq{x21}
C_0=1+k+\gamma_0
\eeq
and 
\beq{x22}
C_n=2n+1+k+\gamma_n-{n(n+k)\over C_{n-1}},~~n=1,2,\ldots 
\eeq
by the same row and column argument as in the proof of 
Theorem 1.  The case $k=1$ is
taken care of by the argument in Theorem 1 since the eigenvalues $\lambda_n$ 
form a decreasing sequence.   In order to prove the
positivity of $C_n$ for general values of $k$ we observe, 
using Eq.\rf{eqm},
that
\beq{coeff0}
q_{nn}=(2n+k+1)\frac{\lambda_{n+1}-\lambda_{n+k+1}}{\lambda_{n}-\lambda_{n+k}}
+n\frac{\Delta\lambda_{n}-\Delta\lambda_{n+k+1}}{\lambda_{n}-\lambda_{n+k}}+(n+k)\frac{\Delta\lambda_{n+1}-\Delta\lambda_{n+k}}{\lambda_{n}-\lambda_{n+k}}\;,
\eeq
where
$
\Delta\lambda_n=\lambda_{n-1}-\lambda_n\;,\quad n\geq 1.
$
 Furthermore, 
\beq{x24}
\Delta\lambda_n>\Delta\lambda_{n+1}
\eeq
for $n\geq 1$, since $V'(\lambda_n)>0$ for $n\geq 1$ in the case at hand, $N=0$.
We have
\beq{x25}
C_0=(k+1){\lambda_1-\lambda_{k+1}\over \lambda_0-\lambda_k}+
k{\Delta\lambda_1-\Delta\lambda_k\over \lambda_0-\lambda_k}
\eeq
and therefore, since $\Delta\lambda_1\geq \Delta\lambda_k$,
\beq{x26}
C_0\geq (k+1){\lambda_1-\lambda_{k+1}\over \lambda_0-\lambda_k}.
\eeq
Finally, using Eqs.\ \rf{coeff0} and \rf{x24}, it follows by induction that
\beq{x27}
C_n\geq (n+1+k){\lambda_{n+1}-\lambda_{n+k+1}\over \lambda_n-\lambda_{n+k}}
\eeq
and the proof is complete.

\medskip
\noindent
{\bf Proof of Theorem 4.}
We only need to consider  $N\geq 1$.
As explained in the proof of Theorem 3 it suffices to show that 
there exists a number $\theta_c$ such that the
$C_i$'s, defined inductively by Eqs.\ \rf{x21} and \rf{x22}, are positive for
each value of $k=0,1,2,\ldots$, provided $\theta \geq \theta_c$.
Note that for 
$k=0$ we simply have $\gamma_i=\oh\theta V''(\lambda_i)$.

We begin by discussing the case $k=0$ and choose $\theta_c$ such that
\beq{444}
V''(\lambda_m)\geq 0
\eeq
for $\theta\geq \theta_c$ and $m=0,1,\ldots$.  Then $C_0\geq 1$ and it
follows easily by induction that $C_m\geq m+1$ for $m>0$.

The case $k=1$ follows from the proof of Theorem 2 
since $\{\lambda_n\}$ is by construction monotonically decreasing. 

In general $\{V'(\lambda_n)\}$ is not a positive decreasing sequence for
$n\geq 1$ so the argument used in the proof of Theorem 3 does not generalize
and we will need to use information about the asymptotic behaviour
of the eigenvalues of $\vp_N$ as $\theta\to\infty$.

We begin by analysing the asymptotic beahaviour of the eigenvalues of
$\vp_N$ regarded as functions of $\theta$.   By Theorem 1 we can write
the eigenvalues as
\bea
\lambda_i(\theta ) & = & s-r_i(\theta ), ~~i=0,1,\ldots ,N\\
\lambda_i(\theta ) & = & r_i(\theta ), ~~i=N+1,N+2, \ldots ,
\eea
where $r_i(\theta )\to 0$ as $\theta\to\infty$ for all $i$.  
The potential function
$V$ is assumed to be $C^2$ and $V''(0)>0$, $V''(s)>0$ so the equation of
motion \rf{7} used for $m=0$ implies that
\beq{x30}
r_0(\theta )-r_1(\theta )=-{\theta\over 2}\left[
V''(s)r_0(\theta)+o(r_0(\theta ))\right]
\eeq
which shows that $\theta r_0(\theta )\to 0$ as $\theta\to\infty$.
Repeating this argument for the next values of $m$ we find that
\beq{x31}
\theta r_i(\theta )\to 0,~~i=0,1,\ldots N-1.
\eeq
Using \rf{x31} in the equation of motion for $m=0,1,\ldots ,N-1$ we find
by an analogous argument that
\beq{x32}
\theta ^2 r_i(\theta )\to 0,~~i=0,1,\ldots N-2.
\eeq
Continuing in the same vein we obtain
\beq{x33}
\theta^{N-i}r_i(\theta )\to 0~~{\rm as}~~i=0,1,\ldots N-1. 
\eeq
Using \rf{x33} in Eq.\ \rf{7} with $m=N$ gives
\beq{x34}
\theta V'(\lambda_N(\theta ))\to -2(N+1)s,
\eeq
which implies
\beq{x35}
r_N(\theta )={2(N+1)s\over V''(s)\theta} +o(\theta ^{-1}). 
\eeq
Continuing this argument we find
\beq{x36}
r_{N}(\theta )\sim {d_N\over \theta},\; r_{N-1}(\theta )\sim {d_{N-1}\over 
\theta^2},\ldots ,r_0(\theta)\sim {d_0\over \theta ^{N+1}},
\eeq
where
\beq{x37}
d_N={2(N+1)s\over V''(s)}.
\eeq
We do not need the explicit values of $d_i$ for $i=0,\ldots N-1$.
Using \rf{x36} in Eq.\ \rf{7} with $m=N+1$ yields
\beq{x38}
r_{N+1}(\theta )\sim {d_{N+1}\over \theta},
\eeq
where 
\begin{equation}
 V''(0)d_{N+1}=V''(s)d_N = 2(N+1)s\;.  
\end{equation}
Taking now $m>N+1$ in Eq.\ \rf{7} we find
\beq{x39}
\theta r_i(\theta )\to 0~~{\rm as}~~\theta\to\infty
\eeq
for $i\geq N+2$.  Continuing the analysis in the same fashion as for $i\leq N$
we obtain the bound
\beq{x40}
r_i(\theta )=O(\theta^{N-i})
\eeq
for $i\geq N+2$.
This completes our discussion of the behaviour of the eigenvalues of $\vp_N$ 
for large $\theta$.

We now use the asymptotic behaviour of the $\lambda_i$'s to find the 
asymptotic behaviour of the $\gamma_i$'s.  This is a straightforward
calculation using Eq.\ \rf{em} and Eqs.\ \rf{x36}-\rf{x40}.  The results
can be summarized as follows:
\bea
{\bf  k=2}~~~~~~~~~~~~~~~~~~~&&\nonumber\\
  m\leq N-2 & : & \gamma_m={\theta\over 2}V''(s)+O(1)
\label{67}\\
  m\geq N+1 & : & \gamma_m={\theta\over 2}V''(0)+O(1)\label{68}\\
  m=N-1 & : & \gamma_m=-(N+1)+{(N+2)d_{N+1}+d_N\over s\theta} +
O(\theta ^{-2})\label{69}\\
  m=N & : & \gamma_m=-(N+1)+{Nd_{N}-d_{N+1}\over s\theta} +
O(\theta ^{-2})\label{70}
\eea

%\newpage

\bea
{\bf  k\geq 3}~~~~~~~~~~~~~ && \nonumber\\
  m+k\leq N & : & \gamma_m={\theta\over 2} V''(s)+O(1)\label{71}\\
  m\geq N+1 & : & \gamma_m={\theta\over 2}V''(0)+O(1)
\label{72}\\
  m+k=N+1 & : & \gamma_m=-(N+1)+{(N+1)d_{N}+(N+2)d_{N+1}\over s\theta} +
O(\theta ^{-2})\label{73}\\
  m=N & : & \gamma_m=-(N+1)+{(N+1)d_{N+1}+Nd_{N}\over s\theta} +
O(\theta ^{-2})\label{74}\\
 m+k=N+2 & : & \gamma_m=-{Nd_{N}\delta_{k3}+(N+2)d_{N+1}\over s\theta} +
O(\theta ^{-2})\label{75}\\
  m=N-1 & : & \gamma_m=-{(N+2)d_{N+1}\delta_{k3}+Nd_{N}\over s\theta} +
O(\theta ^{-2})\label{76}\\
\mbox{\rm All other cases} & : & \gamma_m=O(\theta^{-2}).\label{77}
\eea
All the correction terms to the above asymptotic expressions are uniform
in $k$ and $m$ for $\theta\geq \theta_c$ and $\theta_c$ sufficiently large.

We are now ready to show that $C_m>0$ for all $k\geq 2$ 
provided $\theta$ is sufficiently large. First, we note that it is an
immediate consequence of the preceding asymptotic formulae and the recursion
relations \rf{x21} and \rf{x22} that $C_m>0$ for $n\leq N-k$ and
$\theta\geq\theta_c$, if $\theta_c$ is large enough. It is convenient to
separate the discussion of the remaining values of $m$ into two
cases depending on whether $N-k\geq 0$ or not.

{\bf Case I. $N-k\geq 0$.}
By Eqs.\ \rf{67} and \rf{71},
\beq{x50}
C_0=k+1+\gamma_0\geq k+1+{\theta\over 2}V''(s)+O(1).
\eeq
Choosing $\theta_c$ sufficiently large we also have
\beq{x51}
\gamma_0,\ldots ,\gamma_{N-k}>0
\eeq
and by induction
\beq{x52}
C_m\geq m+k+1+\gamma_m\geq {\theta\over 2}V''(s) +O(1)
\eeq
for $m=0,1,\ldots ,N-k$.

{\bf I.a.} Assume first that $k=2$.  Then we find, using the asymptotic 
formulae above,
\beq{x53}
C_{N-1}=N+{(N+2)d_{N+1}-(N-2)d_N\over s\theta}+O(\theta^{-2})
\eeq
and
\beq{x54}
C_N={4d_N+(N^2+3N+4)d_{N+1}\over Ns\theta} +O(\theta^{-2}).
\eeq
Choosing $\theta_c$ large enough $C_{N-1}$ and $C_N$ are positive and
\bea
C_{N+1}& = & 2(N+1)+3+\gamma_{N+1}-{(N+1)(N+3)\over C_N}\nonumber\\
       & \geq & {2\theta V''(0)\over N^2+3N+4}+O(1).
\eea
For $\theta$ sufficiently large $C_{N+1}\geq N+2$ and it follows by 
induction that $C_m\geq m+1$ for $m\geq N+2$ if $\theta_c$ is so
large that $\gamma_m\geq 0$ for $m\geq N+2$.

{\bf I.b.} Assume next that $k=3$.  Then we find
by a calculation similar to the one in I.a:
\bea
C_{N-2} & = & N-1+{3d_N+(N+2)d_{N+1}\over s\theta}+O(\theta^{-2})\\
C_{N-1} & = & N +{3(N+2)(d_N+d_{N+1})\over (N-1)s\theta}
-{Nd_N\over s\theta}+O(\theta^{-2})\\
C_N & = & {(N+1)d_{N+1}-3d_N\over s\theta}+
3\,{(N+2)(N+3)(d_{N+1}+d_N)\over N(N-1)s\theta}+O(\theta^{-2})\\
C_{N+1} & = & {\theta\over 2}{18(N+1)V''(0)\over (N+1)^3+11N+17}+O(1).
\eea
Choosing $\theta_c$ sufficiently large the above coefficients are all positive
and taking $\theta_c$ so large that $C_{N+1}\geq N+2$ and $\gamma_m\geq 0$ for 
$m\geq N+2$ we conclude by induction that all the $C_m$'s are positive.

\medskip
{\bf I.c.} Now we consider the case $k\geq 4$.  The calculation is analogous 
to the one given above for $k=2$ and $k=3$.  We evaluate 
$C_{N+1-k}, C_{N+2-k},\ldots C_{N}$ to order $\theta^{-1}$ and find that
$C_{N+1-i}= N+2-i + O(\theta^{-1})$ for $i=2,\dots,k$
and then  
\bea
C_{N} &\geq & \left( N+1 +
  k\frac{(N+k)\cdots (N+2)}{N\cdots (N+2-k)}\right)
  \frac{d_{N+1}}{s\theta} + O(\theta^{-2})\\
C_{N+1} & \geq & {\theta\over
  2}V''(0)\left(1-(N+1+k)\left(N+1+k\frac{(N+k)\cdots (N+2)}{N\cdots 
  (N+2-k)}\right)^{-1}\right) + O(1)\;. \nonumber
\eea
Noting that the coefficient of $\theta$ in the last expression is positive
we proceed to show by induction as before that $C_m>0$ for all $m$ provided
$\theta_c$ is chosen large enough.

{\bf Case II. $k\geq N+1$.}
Again it is convenient to split the argument into different subcases.

\medskip
{\bf II.a.} If $N+1=k=2$ then from the asymptotic formulae we find
\bea
C_0 & = & 1+{3d_2+d_1\over s\theta}+O(\theta^{-2})\\
C_1 & = & {4d_1+8d_2\over s\theta}+O(\theta^{-2})\\
C_2 &\geq & {\theta\over 4}V''(0)+O(1)
\eea
and the argument can be completed by induction as before, provided 
$\theta_c$ is taken large enough.

{\bf II.b.} In the case $N=1$ and $k\geq 3$ we find
\bea
C_0 & = & k+1-{3d_2\delta_{k3}+d_1\over s\theta} +O(\theta^{-2})\\
C_1 & = & k+\left( 2-{3\delta_{k3}\over 1+k}\right) {d_2\over s\theta }
+{kd_1\over (k+1)s\theta }+O(\theta^{-2}).
\eea
Choosing $\theta_c$ sufficiently large we find that $C_0>0$, 
$C_1\geq 2$ and $\gamma_m>0$ for $m\geq 2$.  It follows as before that 
$C_m\geq m+1$ for $m\geq 2$.

{\bf II.c.} Consider $N+1=k=3$.  
The crucial coefficients in this case are $C_2$
which is of order $\theta^{-1}$ and $C_3$ which diverges at large $\theta$.
We find
\bea
C_2 & = & {33d_3+27d_2\over s\theta} +
O(\theta^{-2})\\
    &\geq & {198\over V''(0)\theta} +O(\theta^{-2})
\eea
and consequently
\beq{x58}
C_3={9 V''(0)\theta\over 22}+O(1).
\eeq
Taking $\theta_c$ large we can now complete the argument by induction as 
before.

{\bf II.d.} The case $N=2$ and $k\geq 4$ is quite similar to II.b.  We omit
the details which are straightforward.

{\bf II.e.} Consider the case $N+1=k\geq 4$.  We calculate the $C_m$
inductively, starting with $C_0$ and keeping terms to order $\theta^{-1}$.
We find eventually
\beq{x59}
C_{N-1}=N+{(N+1)(N+2)\cdots (2N)\over 2\cdot 3\cdots (N-1)}
\left({d_N+d_{N+1}\over s \theta}\right)-{Nd_N\over s\theta}+O(\theta^{-2})
\eeq
and after a short calculation
\beq{x60}
C_N\geq {(N+1)d_{N+1}\over s\theta} \left( 1 +{(N+2)(N+3)\cdots (2N+1)\over 
2\cdot 3\cdots N}\right) + O(\theta^{-2})
\eeq
which implies 
\beq{x61}
C_{N+1}\geq {\theta\over 2}V''(0)\left(1-2\left(1+{(2N+1)!\over N!(N+1)!}
\right)^{-1}\right) + O(1)
\eeq
and allows us to complete the argument by induction
provided $\theta_c$ is large enough.

{\bf II.f.} The remaining cases $N\geq 3$ and $k\geq N+2$ are simpler 
than those discussed above. One finds that none of the $C_m$'s approaches zero
for large $\theta$. We omit the details.

This completes the proof of Theorem 4.

\medskip

We end this section by commenting briefly on how to extend the stability
results to dimensions $d>1$.
Even though the eigenvalues of the rotationally invariant operators 
are degenerate in this case the extension of the formula \rf{x10} for the stability
functional $\Sigma$ is straightforward to derive if the potential $V$
  is analytic in a neighborhood of the  interval $[0,s]$,
  as we are assuming.

If we have a solution  $\vp=\sum\lambda_n P_n$ to Eq.\ \rf{6}, we find
 by the analytic functional calculus that
\bea
\Sigma (\omega) & = & \sum_{n=0}^\infty \left(2n+d+
  \frac{\theta}{2}V''(\lambda_n)\right)\Vert P_n\omega P_n\Vert^2_2\nonumber\\ 
 & + & 2\sum_{m<n}
 \left(n+m+d+\frac{\theta}{2}
 \frac{V'(\lambda_n)-V'(\lambda_m)}{\lambda_n-\lambda_m}\right)
 \Vert P_n\omega P_m\Vert^2_2\nonumber\\
&-& 2\sum_{k=1}^d \sum_{\underline n,\underline m}\sqrt{(n_k+1)(m_k+1)}\langle
\underline n +\delta_k|\omega|\underline m +\delta_k\rangle \langle\underline n|\omega|
\underline m \rangle\;, \label{sigma}
\eea
where, as usual, the $P_n$ are the spectral projections of the number
operator, and the standard harmonic oscillator basis vectors are $|\underline
n\rangle$, where  $\underline n =(n_1,\dots,n_d)$ is a multi-index of
non-negative integers. Furthermore, $\delta_1,\dots,\delta_d$ denotes the standard
orthonormal basis for ${\bf R}^d$. 

We see that $\Sigma$ only couples the matrix elements  of $\omega$ diagonally,
i.e., it suffices to show that $\Sigma(\omega)\geq 0$ for
\beq{yyto}
\langle \underline n|\omega| \underline m \rangle =0\quad\mbox{unless}\;\; \underline n -\underline m
=\pm \underline \ell\;,
\eeq
where $\underline \ell$ is an arbitrary integer multi-index, with
$|\underline\ell|\equiv\ell_1+\cdots+\ell_d\geq 0$.

Consider first the case $|\underline\ell|=0$, in which the second sum on the
right hand side of Eq.\
\rf{sigma} does not contribute. If $V''(\lambda_n)\geq 0$ for all $n$, we have
\beq{yy67}
\Sigma(\omega)\geq \Sigma^0_1(\omega)+\cdots+ \Sigma^0_d(\omega)\;,
\eeq
where 
\bea
 \Sigma^0_k(\omega) &=& \sum_{|\underline n|=|\underline
   m|}(n_k+m_k+1)|\langle\underline n|\omega| \underline m \rangle|^2\nonumber \\
&-& 2\sum_{|\underline n|=|\underline m|}\sqrt{(n_k+1)(m_k+1)}\langle\underline
n +\delta_k|\omega| \underline m+\delta_k \rangle\langle\underline n|\omega|
\underline m \rangle\;.
\eea
The contribution to this expression from any fixed values of $n_i$ and $m_i$,
for $i\neq k$, is a quadratic form in the the matrix elements 
\beq{y6}
\langle\underline n|\omega|\underline m \rangle=
\langle n_1,\dots,m_k+\ell_k,\dots,n_d |\omega| m_1,\dots,m_k,\dots,m_d \rangle\;,
\eeq
that may be assumed to be real. It is a simple matter to verify that this
quadratic form is positive definite. Therefore, so is $\Sigma(\omega)$ for
$|\underline\ell|= 0$, provided the condition $V''(\lambda_n)\geq 0$ holds.

For $|\underline\ell|\neq 0$ the first sum on the right hand side of Eq.\ \rf{sigma} 
does not contribute.  
For the coefficient of $\Vert P_n\omega P_m\Vert^2_2$ in the second sum one
obtains the value
\beq{coeff}
(n+m+d)\frac{\lambda_{n+1}-\lambda_{m+1}}{\lambda_{n}-\lambda_{m}}
+n\frac{\Delta\lambda_{n}-\Delta\lambda_{m+1}}{\lambda_{n}-\lambda_{m}}+m\frac{\Delta\lambda_{n+1}-\Delta\lambda_{m}}{\lambda_{n}-\lambda_{m}}
\eeq
by using Eq.\ \rf{6} in the form
\beq{eqmot}
(n+d)\Delta\lambda_n-n\Delta\lambda_{n-1}=\frac{\theta}{2}V'(\lambda_n)\;,
\eeq
where
$
\Delta\lambda=\lambda_{n+1}-\lambda_n$, $n\geq 1$.
This allows us to write
\beq{6677}
\oh\Sigma(\omega) = \Sigma_1(\omega)+\cdots+ \Sigma_d(\omega)\;,
\eeq
where 
\bea
\Sigma_k(\omega) &=& \sum_{\underline n=\underline m +\underline\ell}\{(n_k+m_k+1)\frac{\lambda_{n+1}-\lambda_{m+1}}{\lambda_{n}-\lambda_{m}}\nonumber\\
 &+&
n_k\frac{\Delta\lambda_{n}-\Delta\lambda_{m+1}}{\lambda_{n}-\lambda_{m}}+m_k\frac{\Delta\lambda_{n+1}-\Delta\lambda_{m}}{\lambda_{n}-\lambda_{m}}\}|\langle
\underline n|\omega|\underline m\rangle|^2\nonumber\\
&-& 2\sum_{\underline n=\underline m +\underline\ell}\sqrt{(n_k+1)(m_k+1)}\langle\underline
n +\delta_k|\omega| \underline m+\delta_k \rangle\langle\underline n|\omega|
\underline m \rangle\;.
\eea
Considering terms with fixed values of $n_i,m_i,\;i\neq k$, in this expression
one obtains a quadratic form in the matrix elements that can be handled by an 
 analysis similar to the one that was carried out for the case $d=1$. We
do not elaborate further on the general case here but note 
that the analysis of the one-soliton case, $N=0$, of Theorem 3, generalises
immediately to $\Sigma_k$.  This result is obtained by 
observing that the sequence $\{\Delta\lambda_n\}$
is again decreasing in this case as a consequence of Eq.\ \rf{eqmot} since
$V'(\lambda_n)>0$ for $n\geq 1$. Thus, Theorem 3 also holds for $d>1$.

\section{Nonexistence of smooth families}

In \cite{paper} we proved that rotationally symmetric solutions to Eq.\ \rf{6} do
not exist for sufficiently small values of $\theta$. The purpose of this
section is to prove non-existence of smooth families of solutions for small
$\theta$ without assuming rotational symmetry.  By a smooth family
of solutions we mean a mapping 
from an interval $I\subset \bR$ to ${\mathcal H}_{2,2}$,
\beq{n1}
I\ni\theta\mapsto \vp_\theta\in{\mathcal H}_{2,2},
\eeq
which is continuously differentiable in the norm topology of
  ${\mathcal H}_{2,2}$. 

The proof is based on three lemmas below which are most conveniently
established by representing operators by functions via a
quantization map. 
The Weyl or Weyl-Wigner quantization is perhaps the best known quantization
map.  It
can be defined as the mapping $\pi_W$ which to a function $f(x,p)$ of $2d$
variables, $x,p\in\bR ^d$, 
 associates an operator $\pi_W(f)$ on $L^2(\bR^d)$ whose kernel
$K_W(f)$ is given by
\beq{n2}
K_W(f)(x,y) = (2\pi)^{-d}\int_{\bR^{d}} f\left(\frac{x+y}{2},\; p\right)\; 
e^{i (x-y)\cdot p }dp\;.
\eeq
It is obvious that $\pi_W$ maps Schwartz functions on $\bR^{2d}$ 
bijectively onto
operators whose kernels are Schwartz functions and also maps tempered
distributions onto operators whose kernels are tempered distributions. More
important for the following is the easily verifiable fact that $\pi_W$ maps
$L^2(\bR^{2d})$ isometricaly (up to a factor $(2\pi)^{d/2}$) onto the 
space of Hilbert-Schmidt operators on
$L^2(\bR^d)$,
\beq{n3}
\Vert  \pi_W(f)\Vert  _2^2=\int_{\bR^{2d}}|K_W(f)(x,y)|^2\,dxdy=(2\pi)^{-d}
\int_{\bR^{2d}}|f(x,p)|^2\,dxdp.
\eeq

We shall find it more convenient to use the so 
called Kohn-Nirenberg
quantization $\pi$ for which the kernel $K(f)$ of $\pi(f)$ is given by
\beq{n4}
K(f)(x,y) = (2\pi)^{-d}\int_{\bR^{d}} f(x, p)\; e^{i (x-y)\cdot p }dp.
\eeq
The quantization map $\pi$ clearly has the same
properties as the ones we described for $\pi_W$ above.
Likewise, the following properties of $\pi$ are shared by $\pi_W$ except for
the last one:
  \begin{itemize}
  \item[(a)] If $\pi(f)$ is of trace class then
\beq{n5}
\Tr\pi(f)=\int_{\bR^d}K(f)(x,x)dx=(2\pi)^{-d}\int_{\bR^{2d}}f(x,p)dxdp\;.
\eeq
 \item[(b)] If $g$ depends only on $x$ we have
\beq{n6}
\pi(g(x))= g(x)\;,
\eeq
where the right hand side is to be interpreted as a multiplication operator.
 \item [(c)] If $h$ depends only on $p$ we have 
\beq{n7}
\pi(h(p))=h(\frac{1}{i}\nabla_x)\;.
\eeq
 \item[(d)] If $g$ and $h$ are as above, then  
\beq{n8}
\pi(g(x)f(x,p)h(p))= g(x)\pi(f)h(\frac{1}{i}\nabla_x)\;.
\eeq
  \end{itemize}
From (b) and (c) it follows that 
\beq{n9}
a_k=\frac{1}{\sqrt 2}(x_k+\partial_{x_k})=
\frac{1}{\sqrt 2}\pi(x_k+ip_k)
\eeq
and
\beq{n10}
a_k^*=\frac{1}{\sqrt 2}(x_k-\partial_{x_k})=\frac{1}{\sqrt 2}\pi(x_k-ip_k)\;.
\eeq
From the definition of $\pi$ one then obtains
\beq{n11}
[a_k,\pi(f)]=\frac{1}{\sqrt 2}\pi(\partial_{x_k}f+i\partial_{p_k}f)
\eeq
and
\beq{n12}
[a_k^*,\pi(f)]=\frac{-1}{\sqrt 2}\pi(\partial_{x_k}f-i\partial_{p_k}f)\;.
\eeq
Consequently,
\beq{laplace}
2 \sum_k [a_k^*, [a_k,\pi(f)]]=\pi(\Delta f)\;,
\eeq
 where $\Delta$ is the Laplace operator on $\bR^{2d}$, and the
 (complexification of) the space ${\mathcal
   D}$ introduced in Section 2 is just the image under $\pi$ of the domain of
 definition of the self-adjoint operator $\Delta$. Notice, however, that
 contrary to $\pi_W$ the quantization map $\pi$ does not generally map
 real-valued functions to  self-adjoint operators.

There is to our knowledge no known simple characterisation of the subspace of
$L^2(\bR^{2d})$ consisting of functions $f$ such that $\pi(f)$ is of trace
class. We shall need the following result, depending crucially on property
(d) above, concerning such functions.  Here 
$\Vert  \cdot\Vert  _1$ denotes the
standard trace norm.  

\medskip

\noindent
{\bf Lemma 4.} 
{\it Suppose $f$ is a square integrable function such that  $\pi(f)$
is of trace class. 
Then its Fourier transform  ${\mathcal F}(f)$ is bounded and its uniform norm
$\Vert  {\mathcal F}(f)\Vert  _\infty$ satisfies the inequality
\beq{n15}
\Vert  {\mathcal F}(f)\Vert  _\infty \leq \Vert  \pi (f)\Vert  _1.
\eeq
}
  
\medskip
\noindent
{\bf Proof.} 
First, note that $\pi(e^{-i\xi\cdot x})=e^{-i\xi \cdot x}$ and
$\pi(e^{-ip\cdot\eta})=e^{-\eta \cdot\nabla_x}$ are unitary operators. Hence,
\beq{n16}
\pi(e^{-i\xi\cdot x} f(x,p) e^{-ip\cdot\eta})= e^{-i\xi\cdot x} 
\pi(f) e^{-\eta \cdot\nabla_x}
\eeq
is of trace class and using properties (a) and (d) above we have
 \begin{eqnarray}
{\mathcal F} (f) (\xi,\eta) &=& 
\int_{\bR^{2d}} e^{-i\xi \cdot x}f(x, p) e^{-i p\cdot \eta}dx dp
\nonumber\\ &=& 
\Tr \{ \pi (e^{-i\xi \cdot x} f(x,p) e^{-ip\cdot\eta}) \}
=\Tr \{ e^{-i\xi\cdot x} \pi(f) e^{-\eta \cdot\nabla_x}\} \;,
  \end{eqnarray}
and hence
\beq{n18}
|{\mathcal F} (f) (\xi,\eta)|\leq \Tr (|\pi(f)|)=\Vert  \pi(f)\Vert  _1\;, 
\eeq
which proves the assertion.  

\medskip

Using the above result we get the following a priori estimate relating the
Hilbert-Schmidt and trace norms of any solution of Eq.\ \rf{6}.

\medskip

\noindent
{\bf Lemma 5.} {\it There exists a constant
  $C$, depending only on $V$, such that any solution $\vp$  of 
Eq.\ \rf{6} fulfills
\beq{n20}
    \Vert  \vp \Vert  _2 \leq C\theta^\frac{d}{2}\Vert  \vp \Vert  _1.
\eeq
}

\medskip

\noindent
{\bf Proof.} Since both $\vp$ and $V'(\vp)$ are Hilbert-Schmidt
there exist square integrable functions $f$ and $F$ such that
 $\vp = \pi (f)$ and $V' (\vp) =\pi (F)$.  By Eq.\ \rf{laplace} the
 equation of motion \rf{6} may be written as
\beq{n21}
\Delta f + \theta F = 0
\eeq
or, equivalently,
\beq{n22}
{\mathcal F}(f)(\xi,\eta)=\frac{-\theta}{|\xi|^2 + 
|\eta|^2} {\mathcal F}(F)(\xi,\eta).
\eeq
Using Lemma 4 and the fact that for an appropriate constant
$c$,
\beq{n23}
\Vert  {\mathcal F}(F)\Vert  _{L^2}
= (2\pi)^{2d}\Vert  V' (\vp)\Vert  _2\leq c \Vert  \vp \Vert  _2 ,
\eeq
we get
\bea
(2\pi )^d \Vert  \vp \Vert  _2^2 & = & \Vert  {\mathcal F}(f)\Vert  _{L^2}^2\nonumber\\
             & = & \int_{|\xi |^2 +|\eta |^2 \leq
   \delta^2}|{\mathcal F}(f)|^2 \, d\xi d\eta +\int_{|\xi |^2 +|\eta |^2 > 
   \delta^2}|{\mathcal F}(f)|^2\,d\xi d\eta\nonumber\\
             & = &  \int_{|\xi |^2 +|\eta |^2 \leq \delta^2}
|{\mathcal F}(f)|^2\,d\xi d\eta
    +\theta^2 \int_{|\xi |^2 +|\eta |^2 
    >  \delta^{2}}\frac{|{\mathcal F}(F)|^2}{(|\xi |^2 + |\eta |^2)^2}
\,d\xi d\eta
\nonumber\\
             & \leq & \mbox{ const } \delta^{2d} \Vert  {\mathcal F}(f)\Vert  _\infty^2 
  + \frac{\theta^2}{\delta^4 }\Vert  {\mathcal F}(F)\Vert  _{L^2}^2 \nonumber \\
     & \leq & \mbox{ const } \delta^{2d} \Vert  \vp \Vert  _1^2 
   +c \frac{\theta^2}{\delta^4 }\Vert  \vp \Vert  _2^2 
\eea
for some constant $c$.
If we now let $\delta^4 = c \theta^2$, the result follows.

\medskip

Our next goal is to 
obtain a lower bound on the Hilbert-Schmidt norm of solutions to
Eq.\ \rf{6} .  

\medskip

\noindent  
{\bf Lemma 6.} {\it There exists a constant $C'$, depending only on the potential
$V$, such that any non-zero solution $\vp$  of Eq. \rf{6} satisfies
the inequality 
\beq{n24}
 C' \theta^{-\frac{d}{2}} \leq  \Vert  \vp \Vert  _2.
\eeq
}

\medskip

\noindent
{\bf Proof.} 
Let 
$
\vp = \sum_n\lambda_n P_n
$
be the spectral decomposition of $\vp$, and set, for $a>0$,
\beq{n28}
  \vp_{<a}=\sum_{\lambda_n <a} \lambda_nP_n\;\;  \mbox{ and }\;\; 
\vp_{\geq a}=\sum_{\lambda_n \geq a} \lambda_nP_n .
\eeq
By our assumptions about $V$ we can fix $a>0$ and a
constant $c_1$ such that $V' (\vp_{<a} )$ is positive and
\beq{n29}
\Vert  \vp_{<a}\Vert  _1 \leq c_1 \Vert  V' (\vp_{<a} )\Vert  _1\;.
\eeq
Now, using that
$
\Vert  {\vp}\Vert  \leq s $ and $\Tr (V'(\vp ))=0
$
by Lemma 2, we can estimate $\Vert  V'
(\vp_{<a} )\Vert  _1$ as follows:
\beq{n30}
\Vert  V' (\vp_{<a} )\Vert  _1= -\Tr ( V' (\vp_{\geq a} ) )  \leq 
 \Vert  V' (\vp_{\geq a} )\Vert  _1 \leq c_2 \Vert   \vp_{\geq a}   \Vert  _1 
\eeq
for an appropriate constant $c_2$.
Thus,
\beq{n31}
\Vert  \vp_{<a}\Vert  _1 \leq c_3 \Vert   \vp_{\geq a}   \Vert  _1\;, 
\eeq
where $c_3 =c_1 c_2$. From this we deduce 
\bea
  \Vert  \vp \Vert  _1 & = & \Vert  \vp_{<a}\Vert  _1 +  \Vert  \vp_{\geq a}\Vert  _1 \nonumber\\
             &\leq & (1+ c_3)
    \Vert  \vp_{\geq a}\Vert  _1 \nonumber\\
         & \leq & c_4 \Vert  \vp \Vert  _2^2\;,\label{poi}
\eea
where $c_4=(1+c_3)/a$.
Finally, from \rf{poi} and the a priori estimate of Lemma 5, we get
\beq{n32}
  \Vert  \vp \Vert  _1 \leq Cc_4\theta^{{d\over 2}} \Vert  \vp \Vert  _2
   \Vert  \vp \Vert  _1 
\eeq
from which the claimed inequality  follows.

\medskip

We are now in a position to prove the announced non-existence result.

\medskip

\noindent
{\bf Theorem 5.}   {\it Let $V$ be analytic on a neighbourhood of the interval
$[0,s]$.  Suppose 
\beq{n33}
{( a,b ]} \ni  \theta \mapsto \vp_\theta \in {\mathcal H}_{2,2}\;,
\eeq
where $0\leq a< b$, is a smooth map 
such that $\vp_\theta$ is a nonzero 
solution of the equation of motion \rf{6} for each
$\theta \in (a,b)$. Then $a>0$.}

\medskip

\noindent
{\bf Proof.} Since $\vp_\theta$ is a solution to Eq.\ \rf{6} the
derivative of the energy $S(\vp_\theta)$ with respect to $\theta$ 
is given by
\beq{n34}
  \frac{d}{d\theta}  S(\vp_\theta ) = \Tr V(\vp_\theta ).
\eeq
This is easy to prove using the analytic functional calculus.
Since $V$ is positive definite, it satisfies an estimate of the form
\beq{n35}
V(\vp ) \geq \mbox{ const } \vp^2
\eeq
and hence, by Lemma 6,
\beq{n368}
\frac{d}{d\theta} S(\vp_\theta )  \geq C_V \theta^{-d},
\eeq
where the constant $C_V$ depends only on $V$ (but not on the given family
of solutions). Hence, for $d>1$, the function
\beq{lpo}
\theta \mapsto  S(\vp_\theta )
+ \frac{C_V}{d-1} \theta^{-d+1}
\eeq
is increasing. Now suppose that $a=0$. Then 
\beq{n36}
 S(\vp_\theta ) \leq S(\vp_b )
     + \frac{C_V}{d-1} (b^{-d+1}-\theta^{-d+1})  
\eeq
which contradicts positivity of  $S(\vp_\theta )$
for small $\theta$.

For $d=1$  the expresion $\frac{C_V}{d-1} \theta^{-d+1}$ in \rf{lpo} 
should be replaced by
$-C_V \ln\theta$ and the same conclusion holds.
This proves the theorem.

%\section{Discussion}
%
%At infinite $\theta$ it is natural to interpret the projector $|0\kt\br 0|$
%as a soliton localized at the origin and the coherent state projector
%$|z\kt\br z|$, $a|z\kt =z|z\kt$, as a soliton localized at the point 
%$(x,y)\in\bR^2$ where $z=x+iy$.  In this case one can easily construct 
%multisoliton solutions as explained in \cite{gsm1} and discussed in detail in
%\cite{lindstrom2,gopakumar}.  In this picture the projector
%$|0\kt\br 0|+|1\kt\br 1|$ has an interpretation as two overlapping solitons
%at the origin and similarly one can interpret the projection on the space
%spanned by the first $n$ eigenvectors of the harmomnic oscillator.  The
%solitons $\vp_N$ constructed in this paper are therefore
%overlapping multisolitons (in the case $N\geq 1$)
%at finite $\theta$.  It is an outstanding unsolved problem to construct
%nonoverlapping multisolitons at finite $\theta$.  The functions
%corresponding to such solitons should not be rotationally invariant about any
%point and correspond to operators which are not diagonal in the
%harmonic oscillator basis.  The methods of this paper do not have a
%straightforward generalization to deal with this case.  Indeed, it is not
%obvious that such solitons exist.  The perturbative calculations of 
%\cite{lindstrom2,gopakumar} indicate that there is an attractive potential
%between solitons at finite $\theta$ and there might not be any solution to
%the equation of motion except rotationally inariant ones.

\bigskip

\noindent
{\bf Acknowledgements. } 
The work of B.~D.\ is supported in
part by MatPhySto funded by the Danish National Research Foundation.  
This research was partly supported by TMR grant no. HPRN-CT-1999-00161.
T.~J. is indebted to the Niels Bohr Institute and the CERN theory division
for hospitality.

\end{document}